# Suspicious Transactions in Smart Spaces

Mayra Samaniego
University of Saskatchewan
mayra.samaniego@usask.ca

Cristian Espana
University of Saskatchewan
cristian.espana@usask.ca

Ralph Deters
University of Saskatchewan
deters@cs.usask.ca

## Abstract

*IoT systems have enabled ubiquitous communication in physical spaces, making them smart Nowadays, there is an emerging concern about evaluating suspicious transactions in smart spaces. Suspicious transactions might have a logical structure, but they are not correct under the present contextual information of smart spaces. This research reviews suspicious transactions in smart spaces and evaluates the characteristics of blockchain technology to manage them. Additionally, this research presents a blockchain-based system model with the novel idea of iContracts (interactive contracts) to enable contextual evaluation through proof-of-provenance to detect suspicious transactions in smart spaces.*

## 1. Introduction

According to Zhang et al. [1], physical spaces become smart when its technological elements can communicate between them and with the Internet to enhance the physical features of that particular space and satisfy the requirements of multiple users (Internet of Things multitenancy [2]).

Smart spaces rely on pervasive Internet of Things (IoT) systems [3][4] to get data from the environment and provide accurate services [5]. The more pervasiveness, the more autonomous smart spaces become and the more personalized services multiple users receive [6]. According to Zhang et al. [1], IoT systems that govern smart spaces "must be context-aware so that they can adapt to rapidly changing conditions." However, most of the IoT systems do not evaluate contextual information before executing transactions, which makes IoT networks vulnerable to attacks, — for instance,  the implantable cardiac devices from St. Jude Hospital that were hacked [7].

This research evaluates the characteristics of blockchain technology to handle the emerging concern of evaluating suspicious transactions in smart spaces. Additionally, this research presents a system model to help to identify suspicious transactions in smart spaces by evaluating contextual information and proof of provenance. This system model introduces iContracts (interactive contracts).

The rest of the paper is organized as follows. Section 2 presents smart spaces, their architectures and security concerns. Section 3 introduces suspicious transactions in smart spaces. Section 4 introduces blockchain for smart spaces. Section 5 explains the proposed system model. Section 6 presents preliminary evaluations. Finally, section 7 presents conclusions and future work.

## 2. Smart Spaces

According to Baladin and Waris [8], smart spaces are common spaces that have capabilities to get data from the environment and apply knowledge to fulfill requirements of mobility, distribution and context awareness of its inhabitants, without them to be required to do any particular task.

The advent of Internet of Things (IoT) technology has allowed the development of ubiquitous smart spaces [9][10], which include devices with heterogeneous computing capabilities [11], processing power, memory and communication protocols to satisfy those requirements.

The design of smart spaces is dictated by their mission, location and specific requirements for inhabitants [8]. Traditionally, architectures for IoT smart spaces integrate the Cloud as the provider of virtually unlimited computing resources to process the vast amounts of sensed data and manage the connected devices to satisfy a possible scaling demand [12][13]. This architecture is mostly used in healthcare; for instance, Doukas et al. [14] combine IoT and Cloud computing to manage wearable sensors. Also, Tyagi et al. [15] present and IoT-cloud framework to manage the transferal of patients' health information. Cloud-based smart spaces have to deal with security, latency and traffic caused by the communication, data processing, and access control management of the volume and velocity of data obtained from different geographical locations [13][16][17]. Sometimes, cloud-based smart spaces lack of security at the constrained level of sensors and actuators. This issue is mainly because cloud-based systems do not interact with IoT networks directly but through a middleware [12]. This lack of security at the constrained level is the perfect opportunity for hackers to penetrate our smart spaces (e.g., [7][18][19][20]).

Researchers have presented an alternative architecture to interact with IoT networks directly, which switch the focus to devices [21]. This architecture focuses on the enhancement of sensors and actuators, providing computing resources closer to them. This approach lowers latency and traffic in communication, data processing, and access control management — for instance, enchanted objects customized for specific users [22]. Things-centric smart spaces have to deal with emerging security issues regarding communication, data processing, and access control management at the constrained level where services directly access physical devices [16]. For instance, multitenancy and edge processing [23][2][24].

## 3. Suspicious Transactions in IoT Smart Spaces

Based on the generally accepted architecture of IoT networks [25][22], suspicious transactions can be studied from two perspectives, cloud-centric and things-centric.

The cloud-centric [25] perspective focuses on analyzing communication, data processing, and access control management transactions from services hosted on the Cloud.

The things-centric [22] perspective focuses on analyzing communication, data processing, and access control management transactions in the constrained network and edges. Things-centric suspicious transactions might affect the correct functionality of sensors and actuators, and edge devices that work as bridges.

Table 1 presents a comparative analysis of suspicious transactions in the Cloud and constrained networks.

Table 1: Comparative Analysis of Suspicious Transactions in Smart Spaces

| | **Cloud-Centric** | **Things-Centric** |
|---|---|---|
| **Communication** | - Execution of not registered communication protocols | - Delayed data streaming from sensors and edge devices |
| **Data Processing** | - High latency when processing sensed data on Internet services | - High latency when processing sensed data on edge services hosted at the edge of sensors (e.g. raspberry pi) |
| **Access Control Management** | - Failure of Internet services authentication<br>- No authentication or a global authentication requirement for third parties reading sensed data<br>- Distribution of incorrect views for users<br>- Execution of unauthorized online financial transactions | - Failure of sensors or actuators authentication<br>- Not authentication or a global authentication requirement for third parties executing firmware updates<br>- Distribution of incorrect firmware to devices.<br>- Execution of unauthorized operations among actuators |

In any case, suspicious transactions have to be detected and stopped to avoid compromising the integrity of our smart spaces. According to Lee et al. [26], the contextual scenario in which smart spaces work and data is processed is an emerging security concern.

Additionally, Perera et al. [27] explain that smart spaces should consider surrounding contextual information to guarantee security in rapidly-changing environments. Conceptual and physical conditions can drive this change. The current contextual details of a smart space can determine whether or not a transaction should be considered as suspicious [28][29]. Contextual parameters might include real-time environment data, history data, rules, among others, depending on the application field of the smart space. The following are examples of context-aware IoT systems. Smailagic and Kogan [30] present a location context-aware IoT environment for location sensing, which uses less power and achieves secure privacy. Additionally, Al-Muhtadi et al. [31] present a security scheme that implements a context-aware infrastructure to manage the identification of users and access control to physical resources.

## 4. Blockchain & Smart Spaces

Autonomy and personalization of smart spaces come to the price of giving private data to centralized systems. Smart spaces commonly manage communication, data processing, and access to data through centralized architectures in the cloud. While this approach provides high consistency (CAP theorem [32]), it introduces a security concern as all data and access to data is managed by a unique central authority.

Researchers have integrated blockchain technologies in smart spaces [3][4] to avoid centralization of data. Thus, communication, data processing, and access control management are executed in a distributed approach with high security, reliability, and immutability by blockchain smart contracts without granting all decision power to a single authority [33][34][35].

Even though the immutability of blockchain smart contracts introduces high security and reliability, they might represent a downside when considering security concerns related to rapidly changing contextual information [29][36][37]. Blockchain-based systems have been implemented mostly to manage access control management through smart contracts [38][39][40]. There is still research to do to evaluate blockchain working on rapidly changing environments like smart spaces. It is necessary to build blockchain-based systems that integrate software components to manage rapidly changing contextual information [41][42]. These types of systems would be especially useful nowadays that we are surrounded by IoT devices whose security proxy relies on single keys and passwords. Sometimes, we detect intrusions when it has passed a long time, and our smart space has been compromised.

This research proposes a model that supports blockchain systems by enabling context-awareness and flexibility to detect suspicious transactions under changing contextual information in smart spaces.

## 5. Proposed System Model

We propose a system model that integrates blockchain technology to provide decentralization, high security and reliability of transactions. Additionally, our system model integrates external components to deal with rapidly changing contextual information. We call these external components iContracts (interactive contracts). Unlike traditional blockchain smart contracts, iContracts can be updated at any time, which enables high flexibility that blockchain technologies do not have.

The proposed system model has three components, a private blockchain system to mine transactions, iContracts to evaluate contextual information, and a Transaction Executor to execute transactions over IoT devices. iContracts and Transaction Executor have been designed as REST services, each of them with different URI's as the unique ID. iContracts read information from the private blockchain system through REST microservices implemented with HTTP protocol. iContracts and the Transaction Executor communicate between them through REST APIs implemented with CoAP protocol [43]. The programming language to develop REST microservices, iContracts and Transaction Executor is GoLang [44].

### 5.1. Blockchain System

The blockchain system is the main entry point of the model; it mines all received transactions and passes them to iContracts. The blockchain system is configured as a consortium network to mine transactions through a proof-of-work mechanism. Even though the blockchain network is not public, we consider that the proof-of-work mechanism is necessary to allow the participation of external miners. The blockchain system exposes reading and writing operation through REST API's.

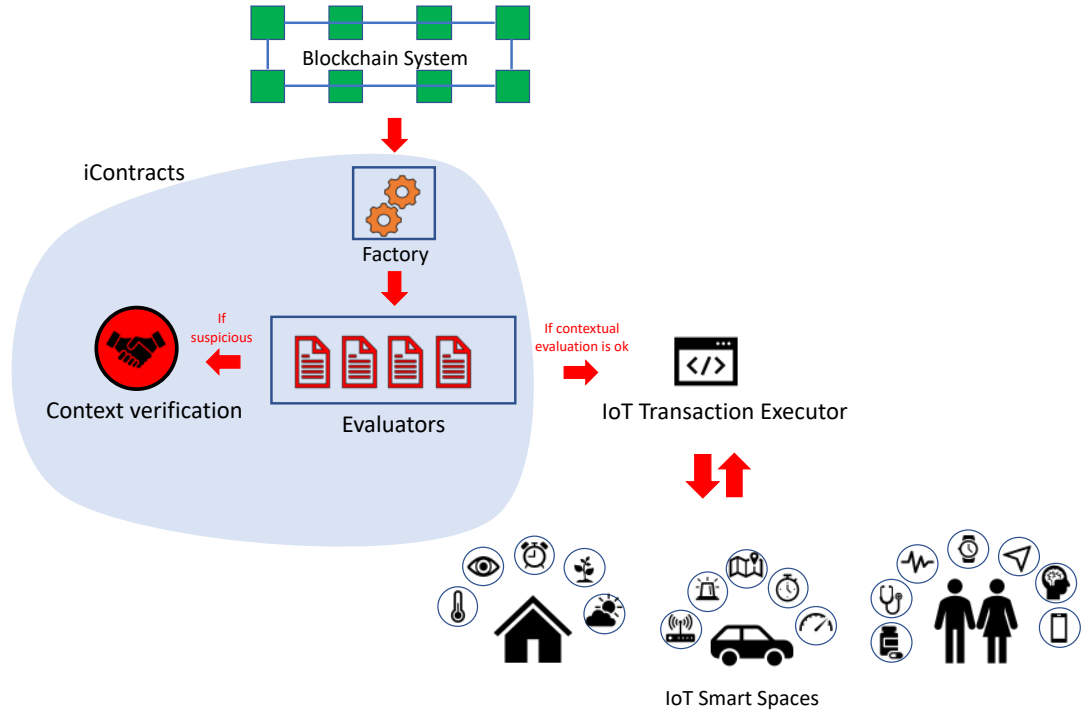

Figure 1. The general architecture of the proposed system model.

### 5.2. iContracts (Interactive Contracts)

iContracts have three levels of abstractions, iContracts factory, iContracts evaluators, and iContracts verifiers. iContracts factory build the contextual information, which is represented by the provenance of transactions of IoT devices read from the blockchain and by the surrounding data got by sensors. iContracts factory builds the skeleton that will allow iContracts evaluators to determine possible suspicious transactions. iContracts factory has the following components:

- Internal URL to receive transactions from the blockchain system, and to interact with other iContracts.
- Conceptual information, the provenance of transactions obtained from the blockchain system.
- Physical Information, surrounding environment data obtained from sensors (e.g. date, temperature, location, among others).

iContracts evaluators encapsulate a set of rules and policies designed as REST services to evaluate contextual information (conceptual and physical information about transactions and IoT devices, respectively) received from iContracts factory. Rules are implemented in the form of If-Else structures, and policies are implemented in the form of triggered actions. If an iContract evaluator detects that a transaction is trying to be executed under and old contextual template or a completely different one, they will consider those transactions as suspicious and stop them. iContracts evaluators mark transactions as suspicious or approved. Approved transactions are sent to the Transaction Executor, and suspicious transactions are sent to iContracts verifiers. iContracts evaluators have the following components:

- Internal URL to receive contextual information from iContracts factory and interact with other iContracts.
- Context-awareness Code, rules/conditions/policies to evaluate contextual information.

iContracts verifiers encapsulate a 2-phase verification process which requires the revoke/approval of users from a trusted device (push notification). All suspicious transactions will be stopped until receiving the commit from the user. iContracts will change the state of suspicious transactions that were accepted by the user to approved and will pass them to the IoT transaction executor. Rejected transactions will be part of the provenance information and will be included in the next round of received mined transactions in iContracts factory.

iContracts can be updated to fulfill the new contextual validation requirements of users. The update of iContracts must be executed by trusted nodes and additionally requires an extra 2-phase commit validation.

### 5.3. Transaction Executor

The transaction executor is the component that finally executes transactions onto physical IoT devices. Only mined transactions that have passed the validation process from iContracts will reach this component. This approach decreases internal latency, which introduces a better performance compared to Cloud-centric IoT [45].

## 6. Preliminary Evaluations

Preliminary evaluations were performed to measure the performance of the two components that support contextual evaluation in smart spaces, iContracts (factory and evaluators). iContracts verifiers could not be evaluated as they depend on the commit action from users. Table 2 presents the characteristic of the nodes that run the iContracts.

We used Ethereum [46] to implement the private blockchain network. Table 2 presents the characteristic of the nodes that run the Ethereum network.

Table 2. Specification of the nodes that run the Ethereum blockchain network and execute iContracts

| Hardware | Details |
|---|---|
| Operating System | Linux Debian 9.8 |
| CPU | Intel(R) Core (TM) i7-6700 CPU @ 3.40GHz |
| RAM | 16 GB |

We emulated IoT temperature sensors as Golang routines deployed on Edison Arduino boards (figure 2). Table 3 shows the specifications of the hardware. We emulated provenance of transactions (everyday transactions) that includes only reading transactions over IoT devices. The reading operations involve sensing the current temperature in Celsius, and the updated configuration is intended to change it to Fahrenheit. The new transaction to be executed is the updating of the configuration to perform those reading operations.

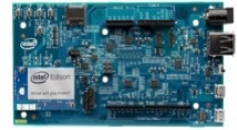

Figure 2. Intel Edison System on a Chip plugged on an Arduino Board [47]

Table 3. Specifications of Edison SoC [47]

| Edison System on a Chip | |
|---|---|
| Operating System | Linux Yocto |
| CPU | 500 MHz dual-core, dual threaded Intel Atom and a 100 MHz 32-bit Intel Quark microcontroller |
| RAM | 4GB LPDDR2 SDRAM |

To have information for the iContract factory to build the contextual scenario, we sent 100 reading transactions. After that, we sent the suspicious transaction (configuration updating) randomly. As soon as the proof of provenance of the iContract evaluator detects a transaction that has not been executed before, they mark the transaction as suspicious and send it to get the user approval or rejection.

Figures 3 to 5 show the performance of the iContract factory while collecting the contextual information under different delays of transaction requests (50, 100, and 200 milliseconds). The contextual information includes the provenance information of the 100 transactions divided into groups of 10 and current date time as the physical information. In the graphs, the x-axis represents every provenance input group, and the y-axis represents the time it takes to the iContract factory to build all the contextual information. The figure shows that it takes around 2 seconds to collect the conceptual and physical contextual information. This performance result is directly linked to the number of previously sent transactions (100). It is necessary to perform more experiments to determine if the size of the provenance data affects the performance of the iContract factory.

The delay in requesting transactions does not affect the performance of the context factory.

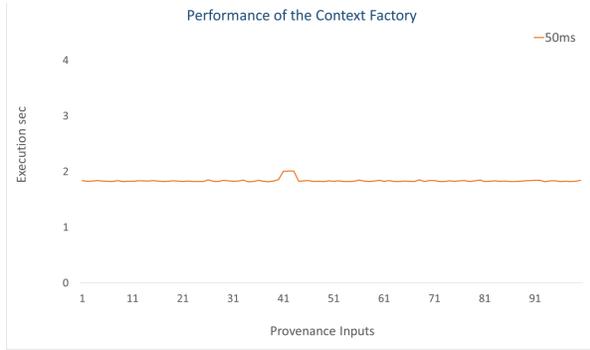

Figure 3. Performance of the iContract factory. 50 milliseconds delay.

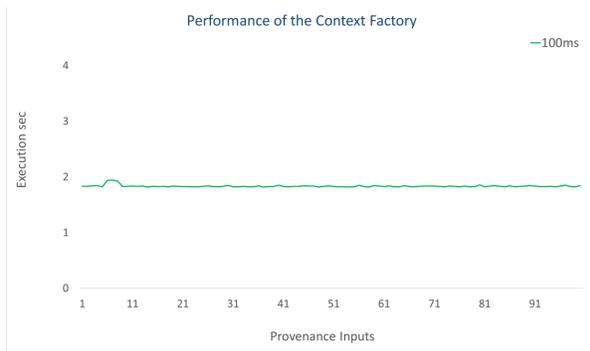

Figure 4. Performance of the iContract factory. 100 milliseconds delay.

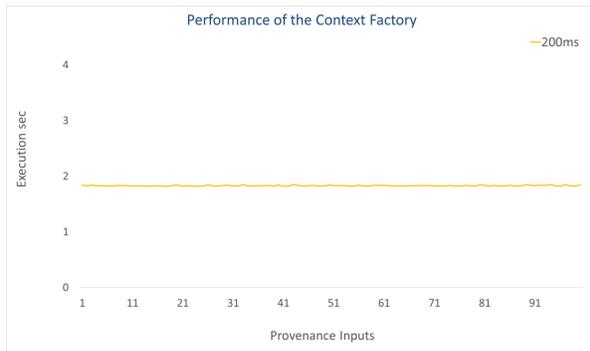

Figure 5. Performance of the iContract factory. 200 milliseconds delay.

Figures 6 to 8 show the performance of the iContract evaluator reading contextual information received from the iContract factory. The results show that in all three scenarios, the response pattern is similar. The iContract evaluator take less than one millisecond to read each contextual input. In average it takes 0.33 milliseconds in all three scenarios. The response times are that low because the iContract evaluator reads every contextual data individually.

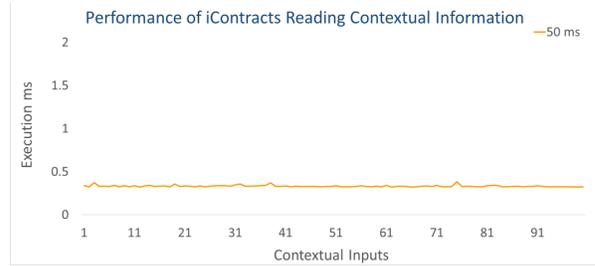

Figure 6. Performance of iContract evaluator reading contextual information. 50 milliseconds delay.

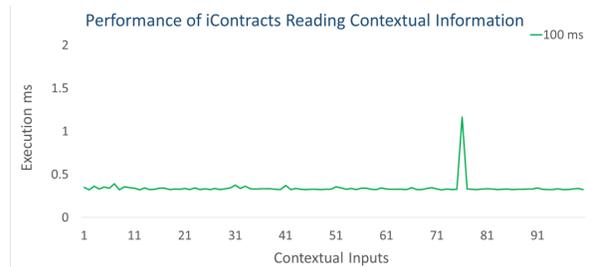

Figure 7. Performance of iContract evaluator reading contextual information. 100 milliseconds delay.

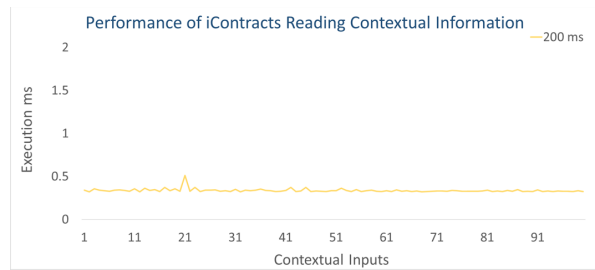

Figure 8. Performance of iContract evaluator reading contextual information. 200 milliseconds delay.

Figures 9 to 11 show the performance of the iContract evaluator analyzing contextual information; this is comparing the provenance and physical data of all 100 contextual inputs. The graph shows that it takes less than 1 second to execute rules and policies contained in the iContract evaluator for each contextual data. This response is because the iContract evaluator works locally. The graphs show some picks. This is because the suspicious transaction is sent randomly, and every time it is detected; then the 2-phase context verification is activated.

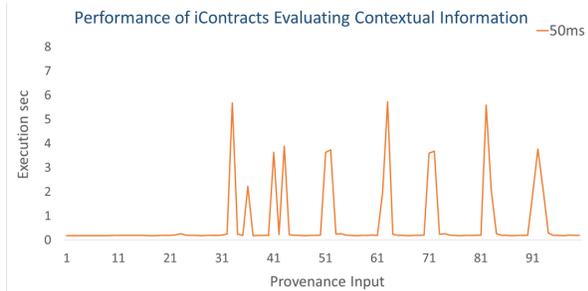

Figure 9. Performance of the iContract evaluator analyzing contextual information. 50 milliseconds delay.

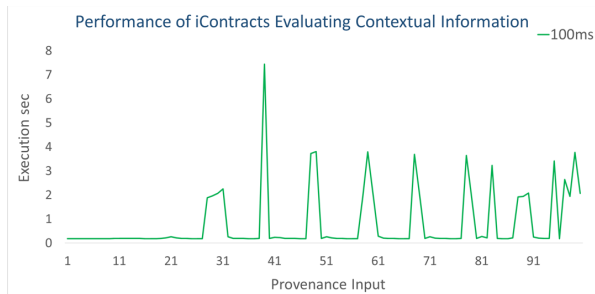

Figure 10. Performance of the iContract evaluator analyzing contextual information. 100 milliseconds delay.

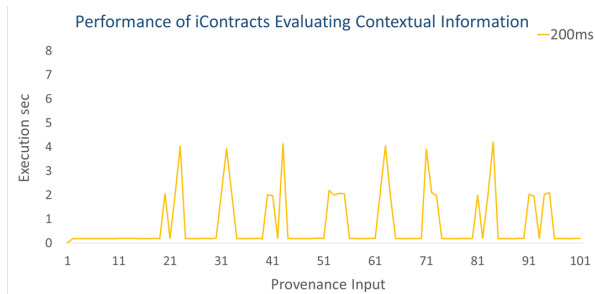

Figure 11. Performance of the iContract evaluator analyzing contextual information. 200 milliseconds delay.

The main goal of this research is to introduce a system that helps to detect the emerging security issue regarding context-awareness and suspicious transactions in smart spaces.

The scope of these experiments does not include the evaluation of the transaction executor and the blockchain system because some research works already evaluate those areas [48][49][50].

For this preliminary evaluation, the iContract verification was designed as a web component, but it can be designed as a mobile component as well. This preliminary evaluation is limited by the number of emulated temperature sensors and the frequency of execution of suspicious transactions. As this is an initial state of the integration of context-aware systems to evaluate suspicious transactions in smart spaces, we have considered that this preliminary evaluation would allow us to get a general vision about how our proposed system behaves in a basic environment. However, more details about contextual information and evaluations will be presented in subsequent research work that integrates extensive experiments environments.

## 7. Conclusions and Future Work

This research presents the novel idea of designing a system model to detect suspicious transactions in smart spaces. This model introduces iContracts (interactive contracts), which are software components that encapsulate information of IoT devices, the provenance of executed transactions, rules and policies to evaluate contextual information and suspicious transactions.

The proposed system model combines blockchain technology with iContracts to enhance security at the constrained level and handle the rapid changing contextual information of smart spaces.

Results of presented experiments may vary due to differences in IoT devices, programming languages, communication protocols, or definitions of different rules/policies in iContracts. As future work, more experiments will be performed to evaluate smart spaces facing different contextual information and working on large test environments.